Evolution of two-step structural phase transition in $Fe_{1+d}Te$ detected by low-temperature x-ray diffraction

Yoshikazu Mizuguchi[1], Kentaro Hamada[1], Kazuki Goto[2], Hiroshi Takatsu[2], Hiroaki Kadowaki[2] and Osuke Miura[1]

1. Department of Electrical and Electronic Engineering, Tokyo Metropolitan University, 1-1 Minami-osawa, Hachioji-shi, Tokyo, 192-0397, Japan
2. Department of Physics, Tokyo Metropolitan University, 1-1 Minami-osawa, Hachioji-shi, Tokyo, 192-0397, Japan

Abstract
The low-temperature crystal structure of $Fe_{1.13}Te$, which exhibits an anomalous two-step magnetic transition, was clarified by the systematic x-ray diffraction measurements. It was found that two-step structural phase transition, tetragonal-orthorhombic-monoclinic, occurred correspondingly to the two-step magnetic transition. The detailed analysis of the profile at 5 K indicated the coexistence of the minor orthorhombic area inside the major monoclinic lattice, which could explain the lower-shift (suppression) of the antiferromagnetic transition temperature in $Fe_{1.13}Te$ and suggest a possibility of superconductivity at the domain boundary.

Introduction

Fe chalcogenide is the ideal system to discuss the mechanism of Fe-based superconductivity, because of the simplest crystal structure and composition among Fe-based superconductors [1,2]. The simplest member, FeSe, crystallizes in the anti-PbO structure, which is composed of only $Fe_2Se_2$ layers stacked along the $c$-axis, and shows superconductivity around 10 K. Furthermore, the transition temperature ($T_c$) exhibits a dramatic enhancement up to 37 K under high pressure around 4-6 GPa [3-6]. The anomalous evolution of high-$T_c$ superconductivity under high pressure can be explained by an optimization of the anion height (Se height) from the Fe plane, which was found to be one of the universal parameters that determine $T_c$ of Fe-based superconductors [7-10]. Very recently we were surprised at the report of an observation of superconductivity above 50 K in a mono-layer FeSe film fabricated on an STO substrate [11].

Another Fe-chalcogenide superconductor is the FeTe-based compound, $Fe_{1+d}Te_{1-x}Se_x$ and $Fe_{1+d}Te_{1-x}S_x$ [12-16]. Although the parent compound $Fe_{1+d}Te$ exhibits antiferromagnetic ordering below ~70 K, superconductivity is induced upon a suppression of the magnetism by partial Se or S substitutions at Te sites. Superconductivity is induced by generating a strain stress in an $Fe_{1+d}Te$ thin film as well [17,18]. However application of external pressure for bulk samples, which generally suppresses antiferromagnetism and induces superconductivity of Fe-based compounds, does not induce superconductivity for $Fe_{1+d}Te$ [19-21]. These facts suggest that superconductivity in the FeTe-based compounds can be obtained only in narrow ranges of both structure and composition (particularly of inter-layer excess Fe composition $d$). To understand the intrinsic nature of FeTe-based superconductivity, clarification of the physical properties of the parent compound $Fe_{1+d}Te$ is much important. Recently structural/magnetic phase diagrams of $Fe_{1+d}Te$ have been reported [22-27]. When the excess Fe concentration is low, tetragonal-monoclinic transition associated with commensurate antiferromagnetic ordering is observed. In contrast, with high excess Fe concentration, tetragonal-orthorhombic transition associated with incommensurate antiferromagnetic ordering is observed. Interestingly, a two-step magnetic transition was observed at an intermediate composition range, $0.11 < d < 0.15$. Synchrotron x-ray and neutron studies suggested an existence of a "mixing phase" in their phase diagrams. The specific heat studies indicated that the higher-temperature transition is a second-order transition, and the lower-temperature transition is a first-order transition [24]. Furthermore, it was suggested that the two magnetic states observed in the mixed phase were competing each other [26]. If the two magnetic states were coexisting in one crystal (lattice) and affecting each other, we expect a strong suppression of the magnetic ordering and appearance of superconductivity at the magnetic/structural domain boundary as filamentary superconductivity observed at a domain boundary of $CaFe_2As_2$ [28]. With this expectation, we performed systematic crystal growth of $Fe_{1+d}Te$ and investigated magnetic properties and low-temperature crystal structure to clarify the nature of the still-unclear "mixed phase" in the phase diagram of $Fe_{1+d}Te$. Here we report an evolution of clear two-step structural phase transition, tetragonal-orthorhombic-monoclinic, in $Fe_{1.13}Te$. On the basis of our experimental results, we suggest an anomalous coexistence of two-types of distortion, monoclinic and orthorhombic, in one crystal. Furthermore, we show a possible superconducting signal observed some samples in susceptibility measurements.

Experimental methods

Single crystals of $Fe_{1+d}Te$ were grown using the self-flux method. Fe powders (99.9 %) and Te grains (99.999 %) were double-sealed in an evacuated quartz tubes, heated to 1050 °C and slowly cooled down to 600 °C with a rate of -4 °C /h. Details in characterization of these crystals are summarized in ref. 27. A polycrystalline sample of $Fe_{1.13}Te$ for low-temperature x-ray diffraction was prepared using a solid-state reaction method. The Fe powders and the Te grains were sealed into an evacuated quartz tube, and heated at 650 °C for 15 h. The product was ground, pelletized, sealed into an evacuate tube and annealed again at 650 °C for 15 h. In this article, we display our results with the nominal compositions. Magnetic susceptibility was measured using a superconducting quantum interference device (SQUID) magnetometer after both zero-field cooling (ZFC) and field cooling (FC). For single crystals, the magnetic fields were applied perpendicularly to the *ab*-plane. X-ray powder-diffraction experiments were carried out using a Rigaku SmartLab powder diffractometer equipped with a $CuK_{\alpha 1}$ monochrometer. The sample was mounted in a closed-cycle He gas refrigerator. For all the x-ray measurements, we firstly cooled the sample space down to the lowest temperature and set it to the target temperature, to avoid an affection of thermal hysteresis of the lattice. In this article, we arbitrarily distinguish mirror indices with the symbols of "T", "O" and "M" for tetragonal, orthorhombic and monoclinic, respectively.

Results and discussion

Figure 1 (a) shows the temperature dependence of magnetic susceptibility after ZFC for $d$ = 0, 0.13, 0.14, 0.2. With increasing $d$, value of susceptibility basically increases because the excess Fe has magnetic moment. For $d$ = 0.13 and 0.14, two-step phase transitions are observed. Figure 1(b) is an enlarged graph around the two-step transition. It is clear that the magnetic transition temperatures for $d$ = 0.13 and 0.14 are shifted to lower temperatures than 70 K in the crystals showing a one-step transition, which is consistent with the previous results [23-27]. To clarify the low-temperature crystal structure for $Fe_{1.13}Te$, we firstly tried the x-ray measurement using the single crystal. However the observed peaks showed drastic broadening after grinding. Although post annealing improved the broadening, the other impurity peaks appeared. Therefore, we performed low-temperature x-ray diffraction using the polycrystalline $Fe_{1.13}Te$ sample. Figure 1(c) shows the temperature dependence of magnetic susceptibility for the $Fe_{1.13}Te$ sample used in the low-temperature x-ray diffraction measurement. The polycrystalline sample also shows a two-step transition as in the single crystals. On cooling, the higher-temperature transition is observed at $T_{N1}$ = 57.9K, and the lower-temperature transition is observed at $T_{N2}$ = 48.8 K, where the transition temperatures are defined as the peak temperature of the differentiation of susceptibility.

The x-ray pattern for the $Fe_{1.13}Te$ polycrystalline sample indicated that the sample was single-phase tetragonal P4/nmm phase. To clarify the evolution of low-temperature structures around the two-step transition, we focused on two reflections of (112) and (200). For this system, the (200) peak splits when the tetragonal lattice distorted to orthorhombic (Pmmn) or monoclinic (P2$_1$/m), and the (112) reflection splits only when the distortion to monoclinic occurred. Figure 2(a) and 2(b) respectively show the (112) and (200) peaks from 100 to 50 K. As shown in Fig. 2(b), a peak broadening of (200)$_T$ starts at ~57 K, and splits into (200)$_O$ and (020)$_O$ at 55 K, which basically corresponds to the middle of the transition at $T_{N1}$. The profile at 50 K is well explained by a superposition of only two split peaks, suggesting that the tetragonal phase is totally distorted. In contrast, the (112) peak does not split at this temperature range in Fig. 2(a). In fact, the transition at $T_{N1}$ is associated with the tetragonal-orthorhombic transition which should be accompanied by the appearance of incommensurate antiferromagnetic ordering. Figure 2(c) and 2(d) show the (112) and (200) peaks from 50 to 5 K, respectively. In Fig. 2(c), the (112) peak begins to exhibit a significant broadening at 47 K, and clearly splits into (112)$_M$ and (11-2)$_M$ below 46 K, which almost corresponds to the transition at $T_{N2}$ in Fig. 1(c). This indicates that an orthorhombic-monoclinic transition occurs at $T_{N2}$, and the magnetic states would be the commensurate antiferromagnetic in this monoclinic lattice. The peak shifts of (200) and (020) are correspondingly observed in Fig.2(d) as well. This result is clearly different from that in the previous report [24]. In fact, the crystal structure of $Fe_{1.13}Te$ at the lowest temperature is basically monoclinic. However, in Fig. 2(c), we find the remained peak of $(112)_O$ around $2\theta = 44.3°$. The population of the orthorhombic phase is estimated to be 20 – 30 % by analyzing an integrated intensities of three peaks of $(112)_O$, $(112)_M$ and $(-112)_M$. On the basis of these facts, we can consider two scenarios. The first scenario is that the small orthorhombic phases (minor phase) exist inside the monoclinic lattice, like an isolated "orthorhombic island" in a "monoclinic sea". In this case, we can understand the suppression of the antiferromagnetic transition temperature, because it was suggested that the two magnetic states in $Fe_{1+d}Te$ are competing. The second scenario is that orthorhombic and monoclinic crystals are coexisting in the sample, namely the two-step transition is not intrinsic for this composition and just a mixing of two phases. However, on the basis of a crystallographic stability and experiences of previous works on this system, the $d$ composition of the orthorhombic phase should be higher than that of the monoclinic phase in this scenario. Hence, the antiferromagnetic transition temperature should be around 70 K, as observed in $Fe_{1.2}Te$ or $Fe_{1.1}Te$. However no anomaly is observed above $T_{N1}$ in Fig. 1(c). Therefore, we think the first scenario is more natural to explain both the structural and magnetic properties of $Fe_{1.13}Te$.

On the basis of our results, we finally propose a phase diagram of $Fe_{1+d}Te$ system in Fig. 3(a). The transition temperature was determined by the magnetic susceptibility measurements for the single crystals. With increasing $d$, the antiferromagnetic transition temperature, ~70 K with $d < 0.1$, is suppressed to ~60 K at $d = 0.11$. Near $d = 0.13$, the orthorhombic phase appears below $T_{N1}$, and finally this phase distorts into monoclinic at $T_{N2}$. In this region, the low-temperature structure would contain partial orthorhombic phase as a minority phase inside the major monoclinic phase as depicted in Fig. 3(b). There are coexisting two magnetic orderings and would be affecting each other. With increasing excess Fe, the orthorhombic phase would gradually become dominant, and finally the monoclinic phase disappears and the orthorhombic phase is dominant at low temperature $d > 0.15$.

If the scenario proposed here is correct, we expect that the magnetism is strongly suppressed at the boundary between monoclinic (commensurate antiferromagnetism) and orthorhombic (incommensurate antiferromagnetism) domains. Furthermore, we think superconductivity can appear at the boundary as observed in the magnetic domain boundary in $CaFe_2As_2$ [28]. In fact, we have observed small drops in the magnetic susceptibility measurements for both single and poly crystals of $Fe_{1+d}Te$ with $d < 0.14$. A typical signal is shown in Fig. 3(c). The susceptibility begins to decrease below ~15 K. The possible $T_c$ is very near that observed in the Fe$_{1+d}$Te superconducting thin film and bulk superconductors. Although we have not observed such a signal in resistivity measurement, we will continue to chase the possibility of domain-boundary superconductivity in Fe$_{1+d}$Te.

Conclusion

We have investigated the magnetic and structural properties of Fe$_{1+d}$Te system. The evolution of the low-temperature crystal structure of Fe$_{1.13}$Te, which exhibits an anomalous two-step transition, was clarified by the detailed x-ray diffraction measurements. The tetragonal lattice totally distorts into orthorhombic at $T_{N1}$. Then, with further cooling, this lattice basically distorts into monoclinic while the minor area remains orthorhombic still at the lowest temperature. These results suggest that two distortions (monoclinic and orthorhombic) are coexisting in a single lattice. Because the coexisting two areas possess the mismatch of both magnetism and lattice distortion, we expect that filamentary superconductivity can occur at the domain boundary.

Acknowledgements

The authors would like to thank Dr. T. Machida (Tokyo Univ. of Science), Dr. H. Sakata (Tokyo Univ. of Science), Dr. Y. Nakai (Tokyo Metropolitan Univ.), Dr. Y. Maniwa (Tokyo Metropolitan Univ.), Dr. Y. Takano (National Institute for Materials Science) for useful discussion. This work was partly supported by Grant-in-Aid for Scientific Research (KAKENHI).

References

[1] F. C. Hsu, J. Y. Luo, K. W. The, T. K. Chen, T. W. Huang, P. M. Wu, Y. C. Lee, Y. L. Huang, Y. Y. Chu, D. C. Yan and M. K. Wu, Proc. Nat. Acad. Sci. 105, 14262 (2008).

[2] Y. Mizuguchi and Y. Takano, J. Phys. Soc. Jpn. 79, 102001 (2010).

[3] Y. Mizuguchi, F. Tomioka, S. Tsuda, T. Yamaguchi and Y. Takano, Appl. Phys. Lett. 93, 152505 (2008).

[4] S. Margadonna, Y. Takabayashi, Y. Ohishi, Y. Mizuguchi, Y. Takano, T. Kagayama, T. Nakagawa, M. Takata and K. Prassides, Phys. Rev. B 80, 064506 (2009).

[5] S. Medvedev, T. M. McQueen, I. Trojan, T. Palasyuk, M. I. Eremets, R. J. Cava, S. Naghavi, F. Casper, V. Ksenofontov, G. Wortmann and C. Felser, Nature Mater. 8, 630 (2009).

[6] S. Masaki, H. Kotegawa, Y. Hara, H. Tou, K. Murata, Y. Mizuguchi and Y. Takano, J. Phys. Soc. Jpn. 78, 063704 (2009).

[7] Y. Mizuguchi, Y. Hara, K. Deguchi, S. Tsuda, T. Yamaguchi, K. Takeda, H. Kotegawa, H. Tou and Y. Takano, Supercond. Sci. Technol. 23, 054013 (2010).

[8] K. Kuroki, H. Usui, S. Onari, R, Arita and H. Aoki, Phys. Rev. B 79, 224511 (2009).

[9] C. Y. Moon and H. J. Choi, Phys. Rev. Lett. 104, 057003 (2010).

[10] H. Okabe, N. Takeshita, K. Horigane, T. Muranaka and J. Akimitsu, Phys. Rev. B 81, 205119 (2010).

[11] Q. Y. Wang, Z. Li, W. H. Zhang, Z. C. Zhang, J. S. Zhang, W. Li, H. Ding, Y. B. Ou, P. Deng, K. Chang, J. Wen, C. L. Song, K. He, J. F. Jia, S. H. Ji, Y. Wang, L. Wang, X. Chen, X. Ma and Q. K. Xue, arXiv:1201.5694.

[12] K. W. Yeh, T. W. Huang, Y. L. Huang, T. K. Chen, F. C. Hsu, P. M. Wu, Y. C. Lee, Y. Y. Chu, C. L. Chen, J. Y. Luo, D. C. Yan and M. K. Wu, Europhys. Lett. 84, 37002 (2008).

[18] M. H. Fang, L. Spinu, B. Qian, H.M. Pham, T.J. Liu, E. K. Vehstedt, Y. Liu and Z.Q. Mao, Phys. Rev. B 78, 224503 (2008).

[14] Y. Mizuguchi, F. Tomioka, S. Tsuda, T. Yamaguchi and Y. Takano, J. Phys. Soc. Jpn. 78, 074712 (2009).

[19] Y. Mizuguchi, F. Tomioka, S. Tsuda, T. Yamaguchi and Y. Takano, Appl. Phys. Lett. 94, 012503 (2009).

[16] Y. Mizuguchi, K. Deguchi, Y. Kawasaki, T. Ozaki, M. Nagao, S. Tsuda, T. Yamaguchi and Y. Takano, J. Appl. Phys. 109, 013914 (2011).

[17] Y. Han, W. Y. Li, L. X. Cao, X. Y. Wang, B. Xu, B. R. Zhao, Y. Q. Guo, and J. L. Yang: Phys. Rev. Lett. 104, 017003 (2010).

[18] Y.F. Nie, D. Telesca, J.I. Budnick, B. Sinkovic, R. Ramprasad, B.O. Wells, Journal of

Physics and Chemistry of Solids 72, 426 (2011).

[19] H. Okada, H. Takahashi, Y. Mizuguchi, Y. Takano and H. Takahashi, J. Phys. Soc. Jpn. 78, 083709 (2009).

[20] C. Zhang, W. Yi, L. Sun, X. J. Chen, R. J. Hemley, H. K. Mao, W. Lu, X. Dong, L. Bai, J. Liu, A. F. Moreira Dos Santos, J. J. Molaison, C. A. Tulk, G. Chen, N. Wang and Z. Zhao, Phys. Rev. B 80, 144519 (2009).

[21] Y. Mizuguchi, F. Tomioka, S. Tsuda, T. Yamaguchi and Y. Takano, Physica C 469, 1027 (2009).

[22] W. Bao, Y. Qiu, Q. Huang, M. A. Green, P. Zajdel, M. R. Fitzsimmons, M. Zhernenkov, S. Chang, M. Fang, B. Qian, E. K. Vehstedt, J. Yang, H. M. Pham, L. Spinu and Z. Q. Mao, Phys. Rev. Lett. 102, 247001 (2009).

[23] E. E. Rodriguez, C. Stock, P. Zajdel, K. L. Krycka, C. F. Majkrzak, P. Zavalij and M. A. Green, Phys. Rev. B 84, 064403 (2011).

[24] S. Rößler, Dona Cherian, W. Lorenz, M. Doerr, C. Koz, C. Curfs, Yu. Prots, U. K. Rößler, U. Schwarz, S. Elizabeth and S. Wirth, Phys. Rev. B 84, 174506 (2011).

[25] I. A. Zaliznyak, Z. J. Xu, J. S. Wen, J. M. Tranquada, G. D. Gu, V. Solovyov, V. N. Glazkov, A. I. Zheludev, V. O. Garlea and M. B. Stone, 1108.5968.

[26] D. Parshall, G. Chen, L. Pintschovius, D. Lamago, Th. Wolf, L. Radzihovsky and D. Reznik, arXiv:1111.4236.

[27] Y. Mizuguchi, K. Hamada, O. Miura, arXiv: 1202.0341. (to be published in ISS-2011 proceeding)

[28] H. Xiao, T. Hu, A. P. Dioguardi, N. Roberts-Warren, A. C. Shockley, J. Crocker, Z. Viskadourakis, X. Y. Tee, I. Radulov, C. C. Almasan, N. J. Curro, C. Panagopoulos, arXiv:1107.0904.

Figure captions.

Fig. 1. (a) Temperature dependence of magnetic susceptibility for Fe$_{1+d}$Te single crystals with $d$ = 0, 0.13, 0.14 and 0.2. (b)An enlargement of the temperature dependence of susceptibility for the single crystals at low temperatures. (c) Enlarged graph of the temperature dependence of magnetic susceptibility for Fe$_{1.13}$Te polycrystalline sample. $T_{N1}$ and $T_{N2}$ are the transition temperatures determined from the temperature differentiation of susceptibility.

Fig.2. Temperature dependence of theta-2theta scans of Fe$_{1.13}$Te powder sample.   (a,b): profiles around (112) and (200) peaks at temperatures between 100 K and 50 K. The only peak split of (200) into (200) and (020) indicates a lattice distortion from tetragonal to orthorhombic structures. (c,d): profiles around (112) and (200),(020) peaks below 50K. An additional peak split of (112) into (112) and (11-2) clearly indicates a lattice distortion from orthorhombic to monoclinic structures.

Fig. 3. (a) Structural phase diagram of Fe$_{1+d}$Te. The low-temperature region for $d \sim 0.13$, indicated with a color gradation, is basically monoclinic while the minor orthorhombic area would coexist inside the major monoclinic lattice as shown in (b). (b) Schematic image of coexisting two structural/magnetic domains. (c) Temperature dependence of magnetic susceptibility for Fe$_{1.13}$Te single crystal. The arrow indicates a drop which will be corresponding to domain-boundary superconductivity.

Fig. 1

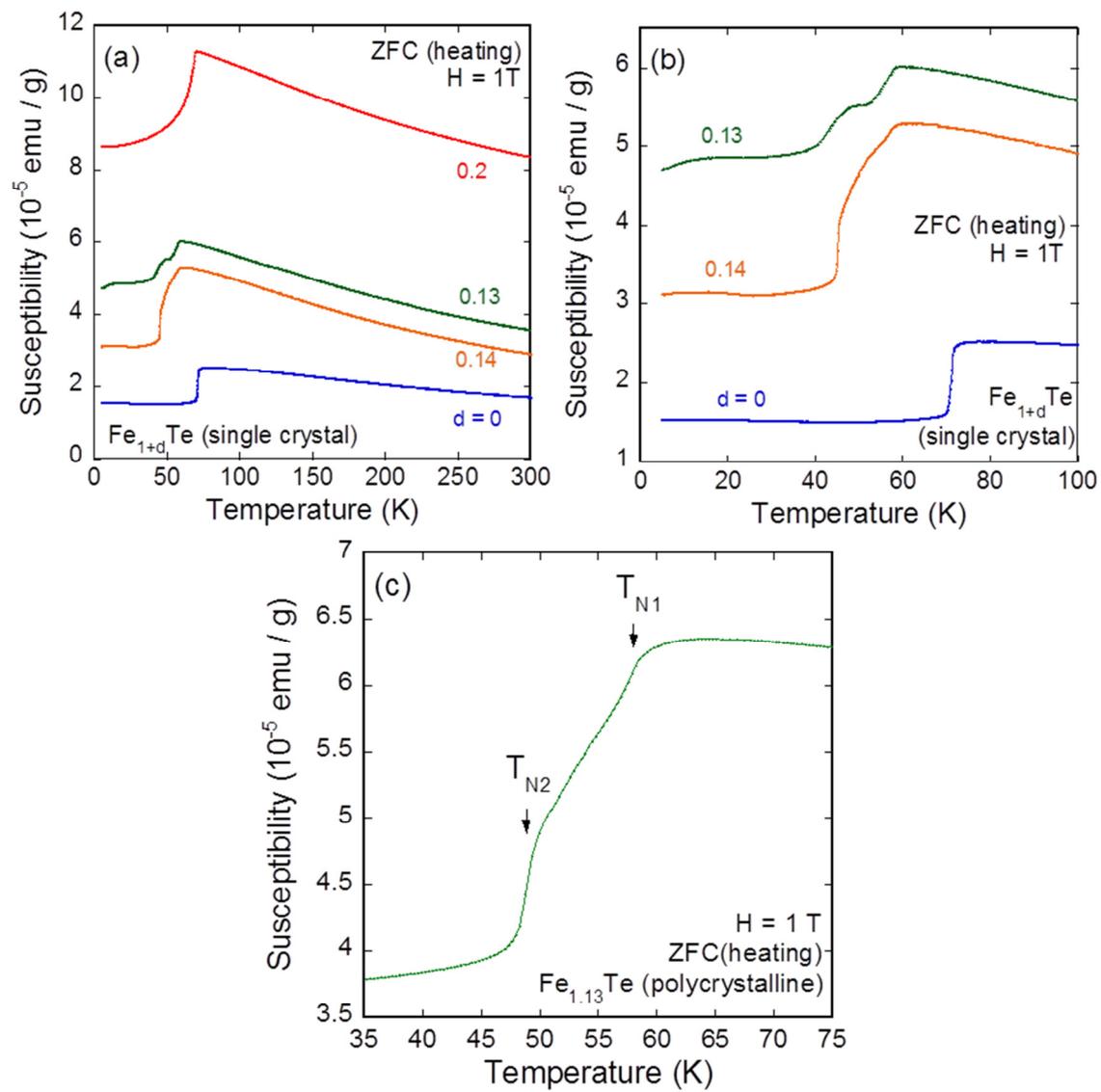

Fig. 2.

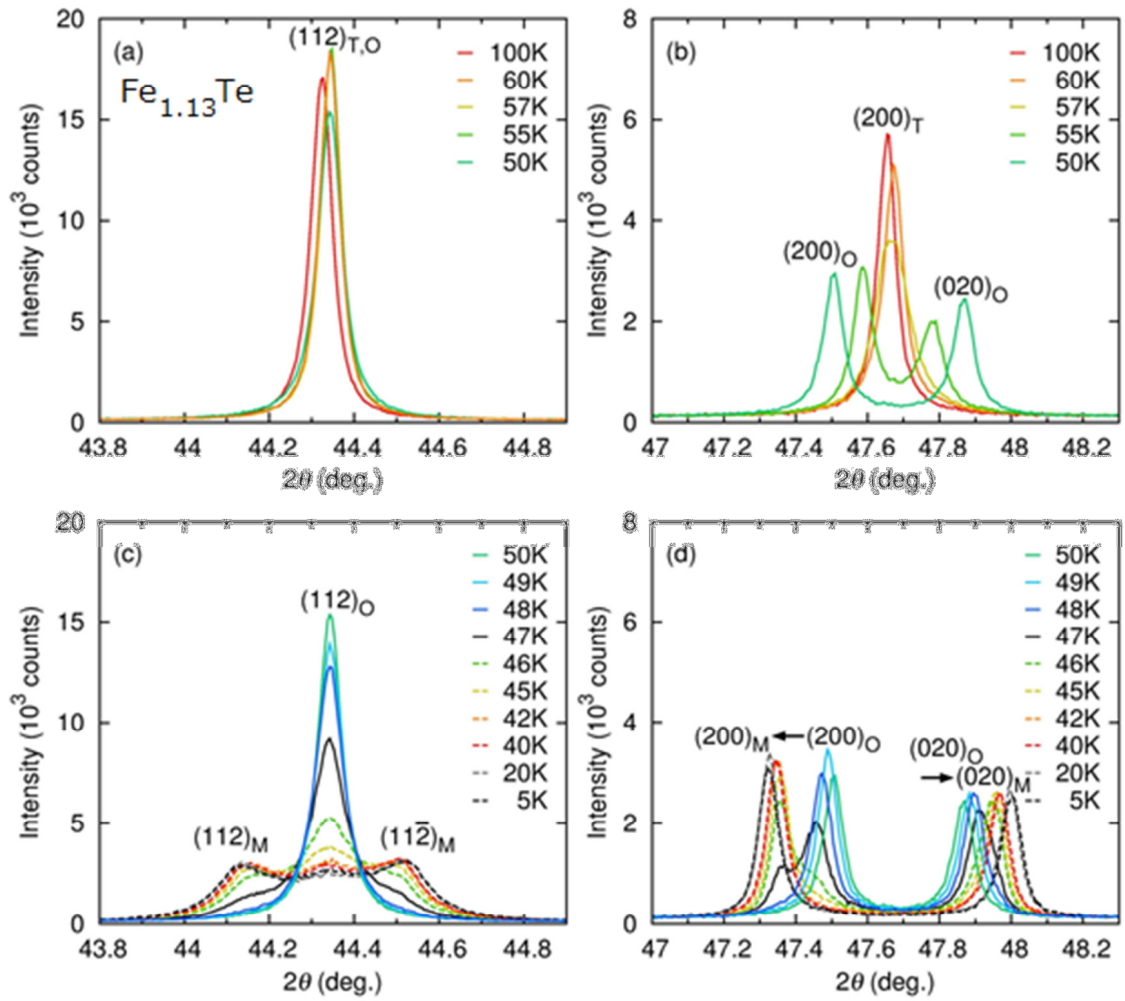

Fig. 3.

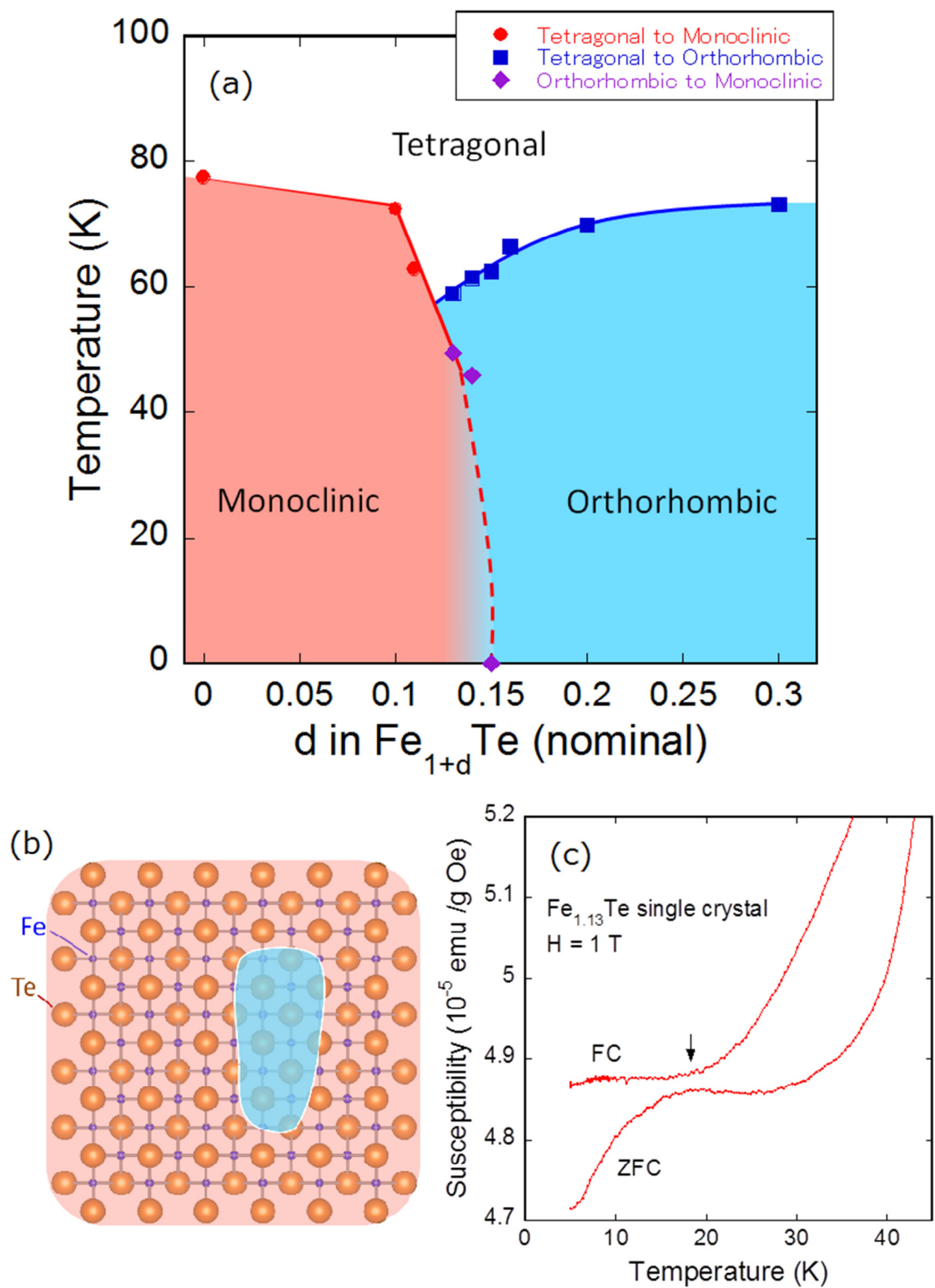